\documentclass[pra,twocolumn,amsmath,amssymb,amsfonts,nofootinbib]{revtex4}

\usepackage[]{graphicx}
\usepackage[]{amsmath}
\usepackage{color}
\usepackage{microtype}

\def\ket#1{| #1 \rangle}
\def\bra#1{\langle #1 |}

\def\kb#1#2{| #1 \rangle\!\langle #2 |}

\def\cC{\mathcal{C}}

\def\cE{\mathcal{E}}

\def\cJ{\mathcal{J}}

\def\cR{\mathcal{R}}

\def\Tr{\mathrm{Tr}}

\def\eq#1{Eq.~\eqref{eq:#1}}

\def\BE{\begin{equation}}
\def\EE{\end{equation}}

\begin{document}

\title{Quantum error correction benchmarks for continuous weak parity measurements}

\author{Gabrielle Denhez}
\affiliation{D\'epartement de Physique, Universit\'e de Sherbrooke, Sherbrooke, Qu\'ebec, Canada J1K 2R1}
\author{Alexandre Blais}
\affiliation{D\'epartement de Physique, Universit\'e de Sherbrooke, Sherbrooke, Qu\'ebec, Canada J1K 2R1}
\author{David Poulin}
\affiliation{D\'epartement de Physique, Universit\'e de Sherbrooke, Sherbrooke, Qu\'ebec, Canada J1K 2R1}

\date{\today}

\begin{abstract}
We present an experimental procedure to determine the usefulness of a measurement scheme for quantum error correction (QEC). A QEC scheme typically requires the ability to prepare entangled states, to carry out multi-qubit measurements, and to perform certain recovery operations conditioned on measurement outcomes. As a consequence, the experimental benchmark of a QEC scheme is a tall order because it requires the conjuncture of many elementary components. Our scheme opens the path to experimental benchmarks of individual components of QEC. Our numerical simulations show that certain parity measurements realized in circuit quantum electrodynamics are on the verge of being useful for QEC.
\end{abstract}

\pacs{}

\maketitle

\section{Introduction}

The physical realization of a quantum information processor represents a formidable experimental and theoretical challenge. The inevitable presence of noise and imperfections in a physical device causes it to deviate from its intended ideal evolution. For any complex information processing task comprising a large number of elementary operations, these deviations can add up to the point of completely invalidating the information content of the processor. Experimental efforts are constantly improving the accuracy in the control of quantum devices and reducing the level of noise, but this progress can only make it so far, and other methods are required to overcome the problem of noise and imperfections.

Quantum error correction (QEC) enables, in principle, the coherent manipulation of quantum information for arbitrarily long times, despite the presence of noise and imperfections in the physical device \cite{Sho95a,Ste96b,Got97a}. This is achieved by encoding the information redundantly in an error-correction code. The information in the code is delocalized, encoded in the many-body correlations of the qubits. As a consequence, it becomes less vulnerable to local noise. To stabilize the information for long times, measurements need to be carried on the redundant systems to detect the presence of possible errors and evacuate entropy from the system. Following these measurements, a recovery operation can be executed to restore the content of the processor to its error-free state. 

 Of course, the building blocks of any error-correcting scheme are themselves subject to faults, and can sometimes introduce more errors in the system than they remove.  The theory of fault-tolerance quantum computation \cite{Sho96a,Kit97c,KLZ98a,AB97a,Pre98a} demonstrates that, provided that the amount of noise in each elementary component is below a certain threshold value, it is possible to use QEC to decrease the level of residual noise after correction below any desired value. 

The exact value of this fault-tolerant error threshold is not know, but it can be estimated numerically for various schemes (e.g. \cite{RH07a,Kni05a}) and theoretical lower bounds can be derived for it (e.g. \cite{R06b,AGP06a}). However, these estimates often require simplifying assumptions and fail to include all the physical features and limitations of a given quantum computer architecture. Theoretical efforts are being deployed to remove these assumptions as much as possible; correlated errors \cite{AGP06a}, slow measurements \cite{DA07a}, and nearest-neighbor interactions \cite{G00a,STD04a,RH07a} are a few examples of limitations  that have recently been incorporated in theoretical estimates of the threshold. It remains very likely that the true threshold deviates significantly from these estimates, and thus experimental investigations of QEC are crucial. 

A handful of experiments \cite{CMP+98a,KLMN01a,CLSB04a,moussa:2011a,zhang:2011a,reed:2012a} have  demonstrated the principles of QEC. Part of the difficulty with the implementation of QEC is that it requires many building blocks, and the failure of one of these could invalidate the entire process. Depending on the details of the scheme, QEC typically requires the preparation of ancilary qubits in a known state, projective measurements, unitary control, classical data processing, etc, and moreover each of these components often needs to be realized in a very short time and with high fidelity.

The goal of this Article is to demonstrate that some of these QEC building blocks can be tested experimentally individually, without necessarily realizing a full QEC cycle. In particular, we propose a scheme to experimentally decide whether a measurement procedure is sufficiently good to perform QEC. We hope that this important simplification will stimulate experimental research in that area, something that could bring new insights to further refine theoretical models. We will pay particular attention to the measurement process which, in many solid-state qubit architecture, can be slow, weak, and noisy.  We will in particular consider weak parity measurements and show that in some conditions such measurements can be successfully used in a QEC scheme to reduce the overall amount of noise. For concreteness, we will illustrate our methods using superconducting qubits in the circuit QED architecture~\cite{blais:2004a,wallraff:2004a}, but the ideas extend to other devices.

Another problem that we address and that is often disregarded is the need for ultra-fast classical data processing to assist the QEC scheme. For instance, the continuous error-correction schemes of Ref.~\cite{ADL02b} is very effective at suppressing errors in theory, but the data processing it entails would require a speed-up of current classical data processing of several orders of magnitudes before becoming practical. Fault-tolerant schemes tailored for slow data processing \cite{DA07a}, more cleaver data-processing algorithms \cite{CLG08a}, or even fast FPGA electronics could surmount these roadblocks, but they nevertheless remain an obstacle to current experimental investigations of QEC. We will show that {\em for the purpose of benchmarking}, all data processing can be delayed until after the experiment at no cost. 

This paper is organized as follows. We first present in Sec.~\ref{sec:QEC} an overview of QEC and highlight some difficulties associated to implementation of these ideas with the measurements typically realized in solid-state qubits. In Sec.~\ref{sec:benchmarking}, we propose a figure of merit that isolates the usefulness of a measurement scheme in a QEC protocol, and propose an experimental procedure to evaluate it. The validity of this experimental procedure is demonstrated in the appendix.  Section~\ref{sec:NumericalResults} presents an numerical application of our method to a recently proposed parity measurement in  circuit quantum electrodynamics architecture~\cite{lalumiere:2010a}. 

\section{Quantum error correction}
\label{sec:QEC}

\subsection{General description}

A QEC cycle usually breaks into three steps, the last one being to some extent optional. In a first step, the system is initialized to a code state $\ket\psi \in \cC$. The quantum code $\cC$ is a subspace of the Hilbert space of $n$ physical qubits that can be specified by a set of operators $S_a$, called stabilizers. The stabilizers represent constraints on the encoded information in the sense that $\ket\psi$ is a code state if and only if it gives the outcome $+1$ when any stabilizer $S_a$ is measured, i.e.~$\cC = \{\ket \psi : S_a \ket\psi = \ket\psi\}$. The dimension of $\cC$ determines the number of encoded qubits. When ${\rm dim}(\cC) = 2^k$, we say that the code encodes $k$ qubits, and that it has a rate $k/n$, the ratio of encoded qubits per physical qubits. For simplicity, we will here limit our discussion to codes encoding a single qubit, $k=1$. 
 
 After the information is encoded, the physical qubits are subject to a noisy evolution. The second step of QEC consists of verifying that the state remains inside the code after this evolution. This is achieved by measuring the stabilizers $S_a$, a measurement outcome that differs from $+1$ signalling the presence of errors. The operators $S_a$ are usually elements of the Pauli group--- i.e., constructed from tensor products of the three Pauli matrices $\sigma_x$, $\sigma_y$, and $\sigma_z$ and the identity---and can therefore only yield $\pm1$ measurement outcomes.  The collection of these $\pm 1$ measurement outcomes, one per stabilizer $S_a$, is called the error syndrome. 
 
 The third and last step consists of reverting, to the greatest possible extent, the effect of the noise. Such a recovery operation can be inferred from a prior statistical description of the noise model---as given, e.g., by a relaxation time $T_1$ and a dephasing time $T_\varphi$---and the additional information obtained from the error syndrome. It is common to assume that the errors $E$ are also elements of the Pauli group, so the statistical evolution of the system is captured by the quantum map $\cE$
\begin{equation}\label{eq:QuantumMapDef}
\cE(\rho) = \sum_E {\rm Pr}(E) E\rho E,
\end{equation}
where ${\rm Pr}(E)$ represents the probability that the Pauli error $E = E^\dag$ has affected the system. In that case, the recovery operation could consist of applying to the system the Pauli operator $E$ that has the largest probability ${\rm Pr}(E)$ and which is consistent with the observed error syndrome. This will be effective since all elements of the Pauli group square to the identity, so applying the same error twice has no net effect on the system. In certain circumstances, this third step can be partly omitted; it is often sufficient to keep in mind that the information store in the encoded qubits has been modified by some known unitary transformation $E$---i.e. perform a basis change---and unnecessary to revert its action. This is sometimes referred to has changing Pauli frame \cite{S03b,Kni05a}. 

\subsection{Difficulties in application to solid-state qubits}

In practice, many of the usual assumptions described above are not met in solid-state devices, and this leads to additional burdens for experimental investigations of QEC. First and foremost, breaking the QEC cycle into discrete steps---a noisy evolution followed by a projective measurement---is not always possible. Measurements of solid-state qubits are often slow, and errors will occur on the system during the course of these measurements.  

In addition to being slow, measurements are noisy. For example, often, the experimental outcome for the measurement of a Pauli operator will be a continuous and noisy signal rather than a binary $\pm 1$ value. This is, for example, the case with an heterodyne measurement in circuit QED~\cite{bianchetti:2009a}. Of course, the noisy signal can be integrated over a period of time and discretized to be treated as binary, but this would entail a loss of precious information about the errors, as we will see below. Measurements can be omitted, but at the price of performing multi-qubit gates (outside the Clifford group) and resetting auxiliary qubits~\cite{reed:2012a}. 

Lastly, the noise affecting solid-state devices (and other qubit architectures) does not correspond to a simple statistical distribution ${\rm Pr}(E)$ over elements of the Pauli group. Instead, we usually model the evolution of these systems by a master equation of the Lindblad form. This more complicated evolution has several consequences. In general, it requires an error-diagnostic protocol far more complex than the one described above for Pauli operators that simply consisted of optimizing ${\rm Pr}(E)$ over a subset of Pauli operators. In addition, the optimal recovery operation found by the error-diagnostic protocol will not be a simple unitary transformation, but instead will be a general quantum map. The physical realization of this map could be beyond today's technologies. Moreover, because the recovery will typically not be a unitary transformation, it cannot be regarded as a basis change and it becomes imperative to implement it physically, an additional experimental burden. One can always pretend that the noise model is a Pauli channel and perform error correction following the usual approach. However, this will inevitably entail a loss of performances, so it does not well reflect the usefulness of the measurement scheme. 

\subsection{Two examples of quantum codes}

We will illustrate our methods to test QEC building blocks using two different error correcting codes. The first one is the bit flip code~\cite{Sho95a}. It uses $n=3$ qubits and has stabilizers $S_1 = \sigma_z\otimes\sigma_z \otimes I$ and $S_2 = I\otimes \sigma_z\otimes \sigma_z$. Thus, the code $\cC$ is spanned by the two states $\ket{\bar 0} = \ket{000}$ and $\ket{\bar 1} = \ket{111}$. As its name suggests, this code is effective at correcting bit flip errors that invert the two basis states $\ket 0 \leftrightarrow \ket 1$, corresponding to the quantum operation $\sigma_x$. On an encoded state $\alpha \ket{\bar 0} + \beta \ket{\bar 1}$, flipping the first qubit would lead to a state of the form $\alpha\ket{100} + \beta\ket{011}$, with error syndrome $S_1 = -1$ and $S_2 = +1$. Similarly, flipping the second qubit produces the syndrome $(-1,-1)$ and flipping the third qubit produces $(+1,-1)$. Thus, any single-qubit flip can be uniquely identified and corrected by an additional flip.  

Flipping two qubits leads to the same syndrome as flipping only the complementary qubit. For instance, an error that flips the first two qubits $E = \sigma_x\otimes\sigma_x\otimes I$ leads to the syndrome $(+1,-1)$, and would be mistaken for the error $E' = I\otimes I\otimes \sigma_x$ flipping only the third qubit. Correcting the error $E$ with the misdiagnosed flip $E'$ would have the net effect of flipping the encoded state from $\alpha\ket{\bar 0}  + \beta\ket{\bar 1} $ to $\alpha\ket{\bar 1} + \beta\ket{\bar 0}$. Thus, the code can correct any one-bit flip, but fails at correcting two-bit flip. For a noise model where each qubit is independently flipped with probability $p$, the code would have a failure probability  ${3 \choose 2} p^2$, a net gain when $p < {3 \choose 2}^{-1} = \frac 13$. 

In practice however, bit-flip is not the dominant source of noise in solid-state devices, so the bit-flip code is not very relevant. We nonetheless choose to use this code to illustrate some of our schemes because it is conceptually very simple. Note that in this paper, whenever we use this code, we will artificially change the noise model of the device, adapting it for the purpose of illustration with the bit flip code. 

The second code we study is tailored to deal with relaxation processes~\cite{LNCY97a}, which are more directly relevant for solid-state qubits. Indeed, relaxation is the main error channel in superconducting qubits, and pure dephasing noise is often negligible in comparison~\cite{houck:2008a}.  The code uses $n=4$ qubits and has stabilizers $S_1 = \sigma_z\otimes\sigma_z\otimes I\otimes I$, $S_2 = I\otimes I\otimes \sigma_z\otimes\sigma_z$, and $S_3 = \sigma_x\otimes \sigma_x\otimes \sigma_x\otimes\sigma_x$. Thus, the code $\cC$ is spanned by the two basis states $\ket{\bar 0} = \frac 1{\sqrt 2}(\ket{0000} + \ket{1111})$ and $\ket{\bar 1} = \frac 1{\sqrt 2}(\ket{0011} + \ket{1100})$. Although the workings of this code is more intricate, one can gain some intuition by noting that when the first qubit relaxes, corresponding to the operator $\sigma^-\otimes I\otimes I\otimes I$,  the initial state $\alpha\ket{\bar 0} + \beta\ket{\bar 1}$ maps to
\BE
\alpha \ket{0111} + \beta  \ket{0100},
\label{eq:deph}
\EE
retaining all the information about the encoded state. 

\section{Benchmarking the measurement}
\label{sec:benchmarking}

\subsection{Isolating the measurement}

Stabilizers can always be measured indirectly by performing an appropriate multi-qubit gate, followed by single qubit measurement. Alternatively,  they can be measured directly using some sort of generalized parity measurement. For instance, the parity measurement proposed in~\cite{mao:2004a,trauzettel:2006a,kerckhoff:2009a,lalumiere:2010a} can be used to measure two-qubit stabilizers, which includes all the stabilizers from the examples described above with the exception of $S_3$ in the last example.  While both approaches are viable, our scheme will focus on such direct stabilizer measurements.
 
The question we would like to answer now is: {\em Can a given physical realization of a parity measurements be successfully used in a QEC scheme to reduce the overall amount of noise?} One way to answer this question is to use the parity measurement in an experimental QEC protocol and check that the resulting noise rate is reduced. However, this method would provide an answer that depends on all the components of the QEC protocol. For instance, perfect parity measurements combined with noisy recovery operations could lead to an overall useless scheme. Here we will show how to decouple these questions and propose a method to evaluate the quality of the measurement alone. 

To decouple the measurement process from the other components of the QEC scheme, it is useful to consider the following two gedanken experiments.  In the first experiment, the system is prepared in a \emph{random} unknown code state $\ket\psi$, and subject to noisy evolution for a time $T$, resulting is a mixed state $\rho = \cE(\Psi)$. Here and throughout, we use capital Greek letters to denote the density matrix corresponding to a pure state $\Psi = \kb\psi\psi$. $\cE$ is the completely positive trace-preserving (CPTP) map that describes the evolution of the system for time $T$. After this time, the optimal recovery operation is performed on $\rho$ to maximize the fidelity to the initial state. This recovery operation can be anything allowed by the laws of quantum mechanics, including measurements, unitary transformations, coupling to an external system, etc. Thus, it can be described by a CPTP map $\cR$. In other words, one chooses the CPTP map $\cR$ that maximizes the quantity $F_\cR(\psi) = \bra\psi\cR(\rho)\ket\psi = \bra\psi\cR(\cE(\Psi))\ket\psi$, averaged over all code states $\ket\psi$. The average performance of this scheme can be summarized by the average fidelity
\BE
\bar F_1 = \max_\cR \int F_\cR(\psi) d\psi.
\label{eq:F1}
\EE 

In the second thought experiment, the system is again prepared in a random unknown code state $\ket\psi$. Then, continuous weak parity measurements of the stabilizers $\{S_i\}$ are performed for a time $T$ to stabilize the encoded states. The outcome of these measurements is a collection of  continuous current $\cJ(t) = \{J_i(t)\}$, one for each measured stabilizer generator. Of course, while these measurements are being performed, the system undergoes its noisy evolution and suffers from measurement back-action. The combined effect of the noisy evolution and the measurements results in a state $\rho  = \cE_\cJ(\Psi)$, where $\cE_\cJ$ is a completely positive (CP) map that depends on the observed currents $\cJ$. At time $T$, after the evolution, one can apply a recovery $\cR$ which yields a fidelity to the initial state $F_\cR(\psi,\cJ) = \bra\psi\cR_\cJ(\rho)\ket\psi = \bra\psi\cR_\cJ(\cE_\cJ(\Psi))\ket\psi$. Clearly, the optimal choice of $\cR$ will depend on the observed currents $\cJ$. The average performance of this scheme can be summarized by the average fidelity
\BE
\bar F_2 = \mathbb E \left[  \max_\cR \left(\int F_\cR(\psi,\cJ)  P(\psi |\cJ) d\psi\right) \right]_\cJ 
\label{eq:F2}
\EE 
where $\mathbb E [\cdot ]_\cJ$ stands for the average over the output currents $\cJ$, whose distribution depends on the initial state $\ket\psi$ and can be modeled by the conditional probability $P(\cJ|\psi)$. The probability $P(\psi|\cJ)$ is obtained from $P(\cJ|\psi)$ via Bayes rule. We denote $\cR_\cJ$ the recovery that maximizes the average fidelity given an output current $\cJ$.

Our original question of whether the parity measurement can be used in a QEC scheme to reduce the overall amount of noise can then be answered positively if $\bar F_1 < \bar F_2$. Indeed, the only difference between \eq{F1} and \eq{F2} is that in the latter, the system's dynamics is altered by the presence of measurements and one uses the outcome of these measurements to design the optimal recovery. The presence of the measurement can lower the fidelity because of the measurement back-action, but it can also increase the fidelity because it brings information about the stochastic evolution of the system. Thus, $\bar F_1 <\bar F_2$ means that including the measurement had an overall positive effect.  

\subsection{Optimal recovery}
\label{sec:OptimalRecovery}

In this section, we explain how to mathematically construct the optimal recovery maps $\cR$ needed for the two gedanken experiments described above. We will focus on the second type of experiment, corresponding to $\bar F_2$ and where continuous parity measurements are performed. The case without measurements follows trivially by removing all references to the output currents $\cJ$. The average fidelity  to the initial state $\bar F$ is a good figure of merit because it has a simple interpretation. However, it is mathematically more convenient to use a slightly different figure of merit called the entanglement fidelity $F_e$ defined as follows. Imagine preparing the system in a state $\ket\phi = \frac 1{\sqrt 2} (\ket{\bar 0}\otimes \ket 0_R + \ket{\bar 1}\otimes \ket 1_R)$ that is maximally entangled between the code space and some auxiliary reference qubit. Here, the subscript $R$ denotes the reference qubit. The entanglement fidelity describes how well this entangled state was preserved by the combined effect of the noise $\cE_\cJ$ and the recovery $\cR_\cJ$:
\BE
F_e(\cJ) = \bra\phi  \cR_{\cJ}(\cE_{\cJ}(\Phi) )\ket\phi
\EE
where it is understood that both $\cR_\cJ$ and $\cE_\cJ$ act only on the system, i.e. the reference qubit evolves trivially.  The average fidelity and the entanglement fidelity are related by $\bar F = \frac{F_ed + 1}{d+1}$ where $d$ is the dimension of the code space, in the present case $d=2$. 

Thus, maximizing the average fidelity is equivalent to maximizing the entanglement fidelity. This is an important conceptual simplification because it only requires optimizing the fidelity for one given initial state---an entangled state between the system and the reference qubit---and therefore leads to a simple numerical procedure. All we need to do is to numerically integrate the master question for the system including noise, given the initial state $\ket\phi$ and taking into account the observed currents $\cJ$~\cite{wiseman:1993a,lalumiere:2010a}. The outcome of such a simulation is a mixed state which we will denote $\Omega_{\cE_\cJ} = (\cE_\cJ\otimes I)(\Phi)$. For a recovery operation $\cR$, the entanglement fidelity can now be expressed as
\begin{equation}
\begin{split}
F_e &= 
 \bra\phi (\cR\otimes I)(\Omega_{\cE_{ \cJ}}) \ket\phi \\
& = \Tr(\Omega_{\cR^\dagger} \Omega_{\cE_\cJ}),
\end{split}
\end{equation}
where $\Omega_\cR = \cR(\Phi)$.  Thus, the optimal recovery can be found by maximizing $ \Tr(\Omega_{\cR^\dagger} \Omega_{\cE_\cJ})$ over $\Omega_\cR$, subject to the constraint that $\cR$ is a CPTP map, i.e. $\Omega_\cR \geq 0$ and $\Tr_R\Omega_\cR = I$, where $\Tr_R$ stands for the partial trace over the reference qubit. This optimization problem is a semi-definite program \cite{FSW07b}, and can be efficiently solved. 

To summarize, the optimal recovery $\cR_\cJ$ for a given set of output currents $\cJ$, can be found by first integrating the  evolution equation for an input state entangled between the code space and a reference qubit, and then solving a semi-definite program. Both of these can be realized in polynomial time as a function of the dimension of the Hilbert space of the system and the duration $T$ of the experiment.

\subsection{Delayed tomography}
\label{sec:delayed_tomography}

So far, we have described how to compute the optimal recovery operation $\cR$ to revert the effect of some evolution $\cE$, that may or may not contain contributions from parity measurements. This recovery is chosen to maximize the average fidelity over the code space. The usual way to assess this protocol experimentally is to perform quantum process tomography on the map  $\mathbb E [R_\cJ \cE_\cJ]_\cJ$ (or $\cR \cE$ in the absence of measurements), describing the joint effect of the measurement and the recovery. This can be realized experimentally as follows.
\begin{enumerate}
\item Prepare a code state $\ket{\psi_i}$.
\item Let the system evolve according to $\cE_\cJ$.
\item Compute the optimal recovery $\cR_\cJ$.
\item Apply the optimal recovery $\cR_\cJ$ to the system.
\item Measure some observable $A_j$.
\end{enumerate}
Repeating this sequence many times provides an estimate of $\mathbb E[\Tr \{A_j \cR_\cJ(\cE_\cJ(\kb{\psi_i}{\psi_i}))\}]_\cJ$. Varying over the initial state $\ket{\psi_i}$ and final measurement $A_j$ provides complete information about the map $\mathbb E[\cR_\cJ \cE_\cJ]_\cJ$. From this complete description, on can directly compute the average fidelity. 

There are two obvious problems with this approach. First, the optimal recovery $\cR_\cJ$ might be beyond what can be realized with today's technologies, so step 4 cannot be realized. Second, when $\cE_\cJ$ is the evolution resulting from the combined effects of noise and parity measurements, the optimal recovery $\cR_\cJ$ depends on the observed measurement currents $\cJ$.  In that case, the procedure to compute $\cR_\cJ$ described in Sec.~\ref{sec:OptimalRecovery} must be realized in real time. With the integration of the evolution equation of a 3-4 qubit system taking  several seconds, the system will have long completely decohered before the recovery operation can even be applied.

We propose a method to directly estimate the average fidelity that circumvents these problems. The intuitive idea is to invert steps 4 and 5 in the above procedure, i.e. measure a complete set of observables before applying the recovery. So crudely, we can imagine that we are characterizing the state $\cE_\cJ(\Phi)$ {\em before} applying the recovery, and then computing what would have been the resulting fidelity had we applied the recovery. The problem with this crude explanation is that the state $\cE_J(\Phi)$ changes at every run of the experiment because it depends on the observed currents $J_j(t)$. Thus, this naive approach only reveals the average state $\mathbb E[ \cE_J(\Phi) ]_\cJ$, which washes away the information content of $J_j(t)$ that should have been used to determine the optimal recovery.

Nevertheless, the essence of the idea is right and, borrowing some ideas from Refs.~\cite{silva:2011a,FL11a}, we will show that the following scheme can be used to estimate the average fidelity.
\begin{enumerate}
\item Choose $\sigma$ randomly among the Pauli operators and a random $\tau = \pm$. Prepare the $n$ physical qubits in the state $\ket{\bar\psi_\sigma^\tau}$, that is the eigenstate with eigenvalue $\tau$ of the encoded operators $\sigma$, i.e.~a state among the set $\{\ket{\bar 0}, \ket{\bar 1}, \ket{\bar 0} \pm \ket{\bar 1}, \ket{\bar 0} \pm i\ket{\bar 1}\}$.
\item Let the system evolve according to $\cE_\cJ$, and collect the currents $\cJ$.
\item Measure the observable $B_k$ chosen uniformly at random from the set of $n$-qubit Pauli operators, obtaining the result $\nu$.
\item Compute the optimal recovery operation $\cR_\cJ$.
\end{enumerate}
This procedure is repeated $M$ times, and the average entanglement fidelity $\bar F_e = \mathbb E[ F_e(\cJ)]_\cJ$ can be estimated by
\begin{equation}
\frac {4^{n+1} }{M} 2 \sum_{\ell = 1}^M  \tau^{[\ell]}  \nu^{[\ell]} \Tr(\Omega_{\cR_\cJ^\dagger}^{[\ell]} B_k^{[\ell]} \otimes\sigma^{[\ell]}),
\label{eq:F}
\end{equation}
where the superscript $[\ell]$ makes reference to the $\ell$th realization of the procedure. In this altered scheme, the experimental work ends at step 3; step 4 and the computation of fidelity can be delayed as post processing. The validity of \eq F is demonstrated in appendix~\ref{sec:ave_entanglement_fidelity}.

\section{Example: parity measurement in circuit QED}
\label{sec:NumericalResults}

\subsection{Parity measurement}

As a concrete example of QEC measurements benchmark, we focus on the 3 and 4-qubit codes discussed above and consider a realization in circuit QED. These ideas are however general and apply to other systems. We will consider the continuous  two-qubit parity measurement proposed in Ref.~\cite{lalumiere:2010a} for circuit QED. Note that two-qubit parity measurements are sufficient to implement the bit-flip code, but insufficient for the relaxation code that also contains the stabilizer $S_3 =  \sigma_x\otimes \sigma_x\otimes \sigma_x\otimes\sigma_x$. In our case study, this third stabilizer will simply be ignored, only the stabilizers that involve a simple two-qubit parity being measured. Since $S_3$ is responsible for the correction of phase errors that occur on a longer timescale $T_2 = 2T_1$, we expect that measurement of the two-qubit parities alone can lead to an improved error rate, and this is confirmed by our numerical simulation.

Before presenting the numerical results, we first give a brief overview of circuit QED. In this architecture, a superconducting qubit is dipole coupled, with strength $g$, to the zero-point electric field of a transmission-line resonator~\cite{blais:2004a,wallraff:2004a}. In the regime where the detuning $\Delta$ between the qubit transition frequency $\omega_a$ and the resonator frequency $\omega_r$ is large with respect to the coupling strenght, $|\Delta| = |\omega_a-\omega_r| \gg g$, the electric-dipole interaction leads to a dispersive coupling of the qubit and the resonator of the form $\chi a^\dag a \sigma_z$, where $\chi = g^2/\Delta$. This interaction can be interpreted as a qubit-state depend shift of the resonator frequency to $\omega_r + \chi\sigma_z$. As a result, a coherent tone   of frequency $\omega_m \sim \omega_r$ and amplitude $\epsilon_m$ driving the resonator will displace the intra-resonator field, initially assumed to be in the vacuum state, to a qubit-state-dependent coherent state $\ket{\alpha_\sigma}$ of amplitude $\alpha_\sigma = - \epsilon_m/\left[\omega_r + (-1)^\sigma\chi - \omega_m -i \kappa/2 \right]$. In this expression, $\kappa$ is the resonator photon-loss rate and $\sigma = \{0,1\}$ stands for the two qubit states. These two coherent state can be resolved by homodyne detection of the coherent drive transmitted or reflected by the resonator, thereby realizing a measurement of the qubit~\cite{WSB05a}.

A similar situation holds if multiple qubits are dispersively interacting with the resonator. For example, in the presence of two qubits, the resonator frequency is shifted to the two-qubit state dependent value $\omega_r + \chi_1 \sigma_{z}\otimes I + \chi_2 I\otimes\sigma_{z}$, where $\chi_j$ is the dispersive coupling of qubit $j$. This leads to four possible qubit-state dependent coherent states $|\alpha_\sigma\rangle$, where now $\sigma= \{00,01,10,11\}$, and that can be resolved in a homodyne detection~\cite{majer:2007a}.

\begin{figure}[t]
\includegraphics[width=0.80\hsize]{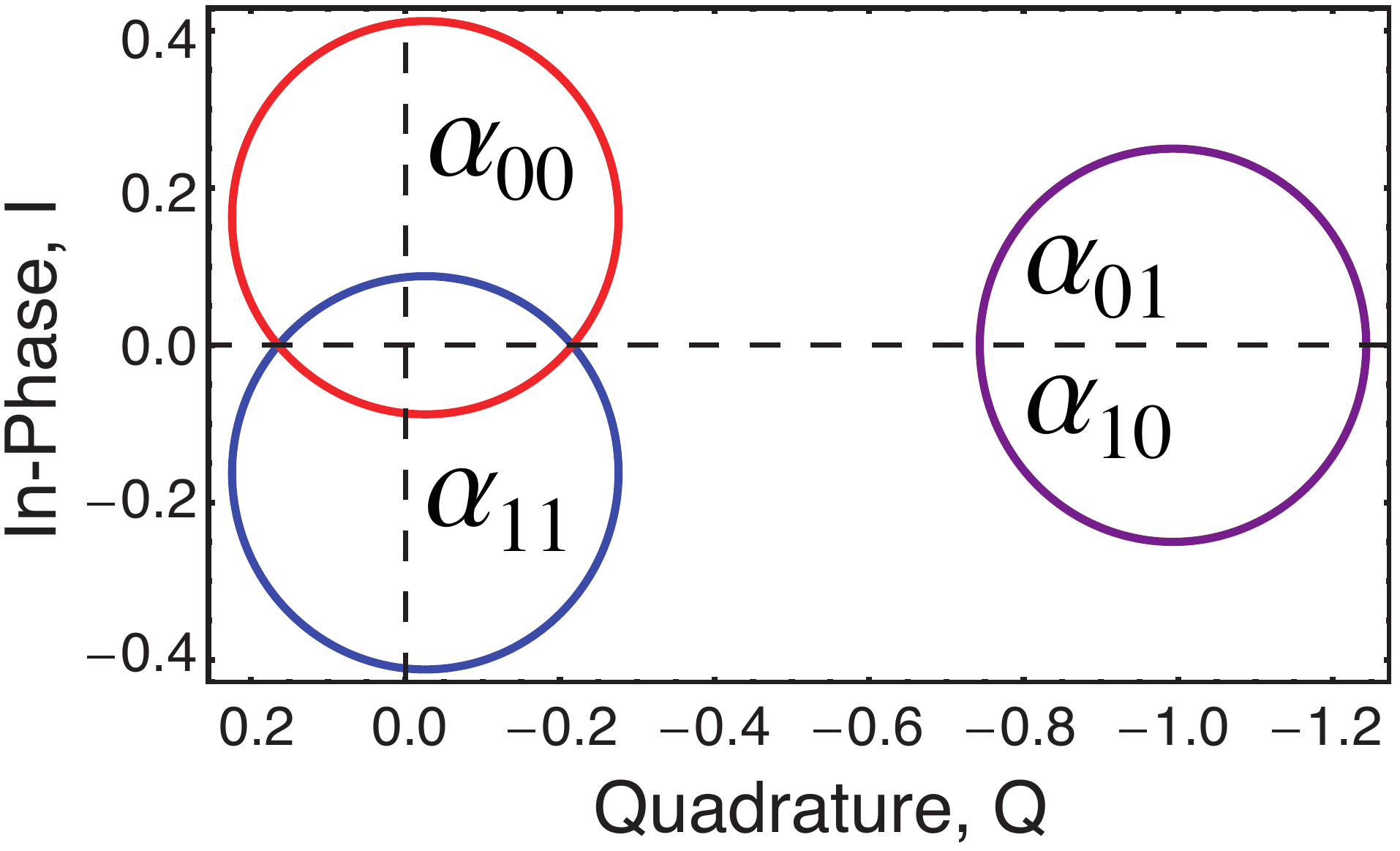}
\caption{(color online)  Phase space illustration of the stationary states $\ket{\alpha_{\sigma\sigma'}}$ for: $g_1=-g_2=-  15\kappa$ and $\chi_{j} \sim 1.5 \kappa$.  The measurement drive is such that $\omega_m = \omega_r$ and $\epsilon_m=\kappa/2$. 
}
\label{fig:CoherentStates}
\end{figure}

As shown in Refs.~\cite{filipp:2009a,lalumiere:2010a}, it is possible to adjust the frequency of the measurement tone, the dispersive couplings and the frequency and phase of the local oscillator used in the homodyne detection such that the measurement only reveals information about the parity of the joint two-qubit states. This is illustrated in Fig.~\ref{fig:CoherentStates} where the parameters have been chosen such that the coherent states corresponding to even qubit parity $\{\ket{00},\ket{11}\}$ overlay, while those corresponding to odd parity $\{\ket{01},\ket{10}\}$ almost do~\footnote{It is not possible for the coherent states corresponding to odd parity to perfectly overlay if the even ones do, and vice-versa. With the parameters chosen for Fig.~\ref{fig:CoherentStates}, this results in measurement-induced dephasing of superpositions of the odd parity states~\cite{lalumiere:2010a}.}. In this situation, homodyne detection of the $Q$ quadrature of the resonator field will yield information about the qubit parity, and nothing about the individual qubit state. In other words, for the settings of Fig.~\ref{fig:CoherentStates}, the measurement operator in circuit QED takes the form $\sigma_z\otimes\sigma_z$, exactly what is required for the stabilizers $S_1$ and $S_2$ of the two quantum error correcting codes discussed above. Moreover, these two stabilizers can be measured simultaneously, for example, by using a two-resonator setup where the pair of qubits to be  jointly measured is fabricated in the same resonator. For the 3-qubit code, this means that one of the qubit must be coupled to two resonators, something that has already been realized experimentally~\cite{johnson:2010b}.

\subsection{Numerical results}

\begin{figure}[t]
\includegraphics[width=0.95\hsize]{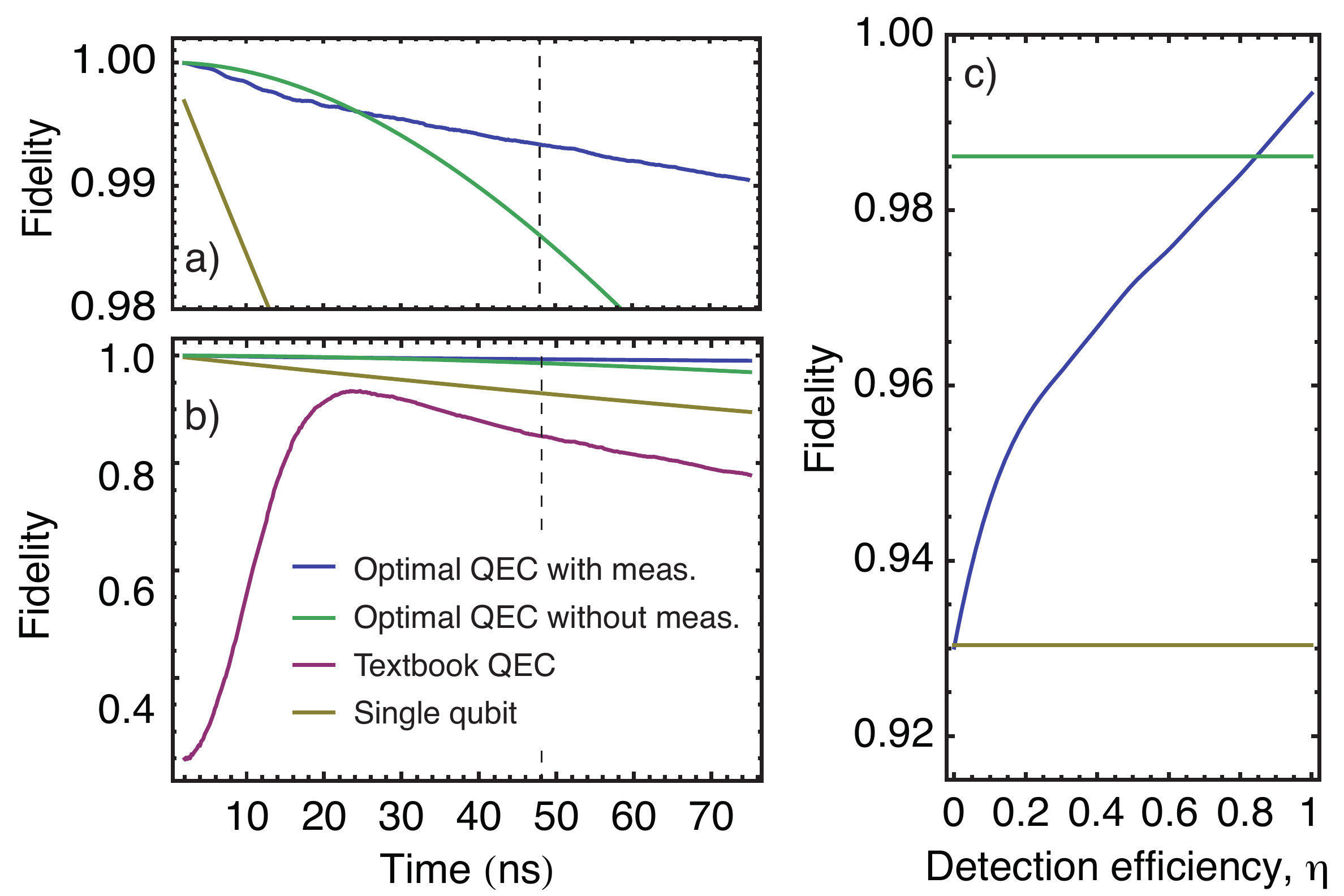}
\caption{(color online) a-b) Fidelity vs time for optimal QEC with (blue line) and without (green line) parity measurements, textbook QEC (purple line) and unencoded single qubit evolution (yellow line).  Panel a) is a zoom of panel b) in the region close to unit fidelity. c) Fidelity vs detection efficiency $\eta$ taken at $t=48$ ns as indicated by the vertical dashed line in panels a) and b). In all cases, the parameters are $\chi/2\pi = 120$ MHz, $\kappa/2\pi = 50$ MHz, $\gamma_x/2\pi = 5$ MHz. When measurement is present, $\eta =1$ [in panels a) and b)] and $\epsilon/2\pi = 40$ MHz. The results where averaged over 8000 trajectories.}
\label{fig:3qc_FvsT}
\end{figure}

Fig.~\ref{fig:3qc_FvsT} presents the fidelity as a function of time for the three-qubit code under continuous measurement of the syndromes $S_1$ and $S_2$. These results were obtained by numerical integration of the stochastic master equation (SME) found in Eq.~(2) of Ref.~\cite{lalumiere:2010a} and using the parameters indicated in the caption.  This SME describes the evolution of two qubits coupled to the same resonator and under continuous parity measurement realized, as discussed above, by homodyne detection of the signal transmitted through the resonator. Since the 3-qubit code protects against bit flips, and not relaxation, we replaced the qubit damping term in this SME by a term representing symmetric bit flips. We have also dropped pure dephasing since both the 3 and 4 qubit codes do not protect against this noise source (and many superconducting qubits in circuit QED suffer only very weakly from pure phase damping~\cite{houck:2008a}).  These are the only modifications made to the results of Ref.~\cite{lalumiere:2010a}.

In this Figure, the blue and green lines correspond to the fidelity after optimal recovery, respectively in the presence and absence of continuous weak parity measurements. These results are compared to the fidelity of a single unencoded qubit (yellow line). For the chosen parameters, we see that it is always beneficial to encode the information (blue and green line are above the yellow one), but that the benefit of weak measurements depends on the duration of the experiment. Thus, even with optimal recovery, continuous syndrome measurement does not always help. Indeed, below $\sim 25$ ns,  measurement backaction~\cite{lalumiere:2010a} disrupts the encoded state more that the acquired information helps. Beyond that time, the measurement record can be used to improve the optimal recovery. 

The purple line shows the fidelity obtained by a direct application of textbook QEC. The continuous weak stabilizer measurements are performed for a time $t$, after which the output signal are integrated and round-up to a $\pm1$ value. The simplest Pauli operator consistent with this round-up error syndrome is applied to the state. At early times, the fidelity of this scheme approaches 1/4. This is because the integration of the output signal is essentially random, so the recovery will essentially consist in flipping one of the 3 qubits  or none at all, each option having probability $1/4$. Since at early times it is likely that the qubit has not suffered any errors at all, only 1 out of these 4 options is correct, explaining the $1/4$ fidelity. The fidelity reaches a maximum at a subsequent time, when the syndrome measurements reveal useful information about the error. Yet, it does not reach the fidelity under optimal recovery. This is because, as explained above, the optimal recovery can be any CPTP map, while this textbook application is limited to a recovery that flips one of the three qubits. As intuitively expected, the time scale where this maximum occurs corresponds roughly to the time scale where weak measurements provide an advantage under optimal recovery; prior to this time the measurement signal is essentially random.

\begin{figure}[t]
\includegraphics[width=0.8\hsize]{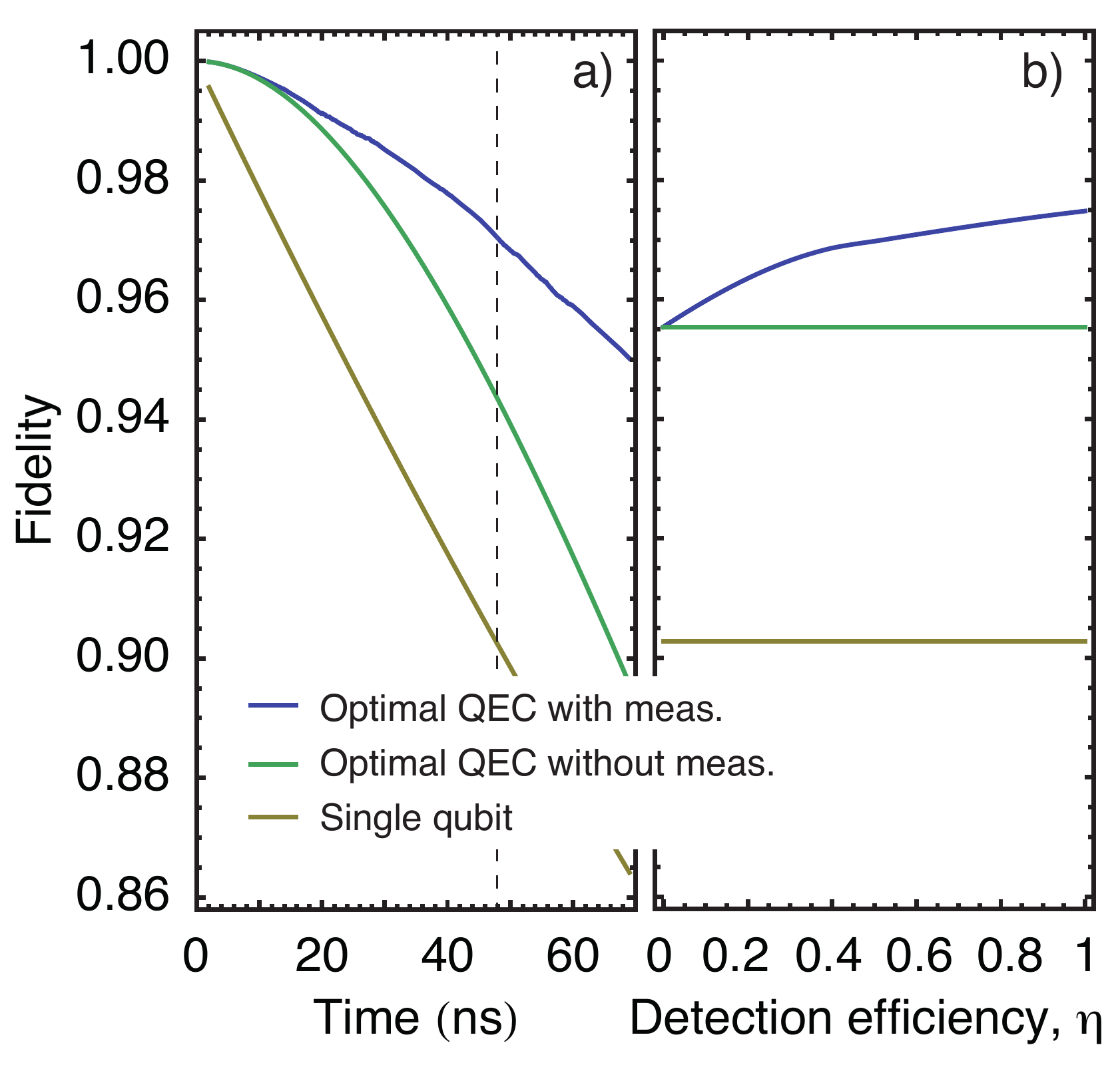}
\caption{(color online) a) Fidelity vs time for optimal QEC with (blue line) and without (green line) parity measurements, and unencoded single qubit case (yellow line). b) Fidelity vs detection efficiency $\eta$ taken at $t=48$ ns as indicated by the vertical dashed line in panel a). The parameters are the same as in Fig.~\ref{fig:3qc_FvsT}.}
\label{fig:4qc_FvsT}
\end{figure}

Fig.~\ref{fig:4qc_FvsT} shows similar results for the four qubit code. Since this code protects against relaxation (and not symmetric bit flips), this is the type of noise that was numerically simulated to generate these results. This situation is therefore closer to the experimental reality. Using the same color code as in Fig.~\ref{fig:3qc_FvsT}, the yellow line corresponds to an unencoded qubit, the green line to optimal recovery in the absence of continuous measurements and, finally, the blue line to optimal recovery with continuous measurement of $S_1$ and $S_2$. Compared to the three-qubit code, the case with measurement always do as good, or better, than without measurement. This is because measurement-induced dephasing acts in the odd-parity subspace~\cite{lalumiere:2010a}. This, combined with the fact that the four-qubit codewords are of even parity and that relaxation never bring a superposition of encoded states to a superposition of odd-parity state\footnote{Note for example that the state in \eq{deph} can be written as $\ket{01}\otimes (\alpha\ket{11} + \beta\ket{00})$, and contains no superpositions of odd parity states that would undergo measurement-induced dephasing.}, yields the enhanced result compare to the three-qubit code.

\subsection{Possible experimental realization}

Is this protocol experimentally realistic? First, we emphasize that, with delayed tomography, the recovery operation need \emph{not} be applied in the laboratory. It is only applied, after the fact, on the tomographically reconstructed state. The most stringent requirement is on the quality of the parity measurements, something that depends on the details of the implementation. In this respect, all the circuit QED parameters that we have chosen are within experimental reach, except one: the detection efficiency $\eta$. This parameter corresponds to the the efficiency with which   photons used for the syndrome measurements are detected. The results presented in Figs.~\ref{fig:3qc_FvsT} and \ref{fig:4qc_FvsT} have been obtained by setting $\eta=1$, above what can now be realized. Indeed, most experiments use cold amplifiers with a rather low detection efficiency $\eta \approx 0.05$. Recent experiments with near quantum limited amplifiers have reached $\eta \approx 1/3$~\cite{BerkeleyCommunication}, and values approaching unity should be realizable in the near future. To evaluate the importance of this parameter, the fidelity, evaluated at $t=48$ ns, is plotted versus $\eta$ in Figs.~\ref{fig:3qc_FvsT}c) and \ref{fig:4qc_FvsT}b). For the three qubit code, because of the measurement-induced dephasing, the weak syndrome measurements do not help until a rather high value of $\eta\sim0.85$.  For the more experimentally relevant case of the four-qubit code, because this backaction does not  damage the encoded state, measurement helps even at low efficiency. 

The two other parameters worth commenting on are the coupling strength $g$ and the relaxation rate $\gamma_1$. Our choice of $\chi/2\pi = 120$ MHz and $\lambda = g/\Delta = 1/10$~\footnote{The value of $\lambda = g/\Delta = 0.1$ is chosen such that the system is in the dispersive regime where the results of Ref.~\cite{lalumiere:2010a} hold. For the same reason, we have also chosen the measurement tone amplitude $\epsilon$ such that the average photon number in the resonator is always smaller than about a tenth of the critical photon number $n_\mathrm{crit} = (\Delta/2g)^2$~\cite{blais:2004a}.} correspond to a qubit-resonator coupling strength of $g/2\pi = 1200$ MHz, a value that is beyond current experiments by about a factor of four~\cite{reed:2012a}. An architecture to realize this type of `ultra-strong' coupling has been theoretically proposed~\cite{bourassa:2009a} and tested experimentally with $g/2\pi \sim 600$MHz~\cite{niemczyk:2010a,forn-diaz:2010a}. The value used here is thus not beyond reach.  Finally, the relaxation rate $\gamma_1/2\pi = 5$ MHz used here is more than an order of magnitude larger than recently achieved~\cite{paik:2011a,rigetti:2012a}. Adjusting this rate to the best current experimental value would relax the constraints on all the other parameters. 

Lastly, our scheme is sensitive to state initialization errors. These are estimated to be at the percent level~\cite{palacios-laloy:2009a} but can be reduced by using measurements to initialize the qubits~\cite{johnson:2012a,riste:2012a}.

\section{Discussion}

The method we have presented offers the possibility to experimentally determine if a measurement scheme is useful for a given QEC protocol, without the need to realize the full QEC protocol. In particular, there is no need in our method to implement the recovery operation on the system and all data analysis are delayed until after the experiment---in particular there is no operation that is realized on the system conditioned on the outcome of previous measurements. 

There are obvious limitations to our scheme. On the one hand, a negative answer does not imply that the measurement scheme is of no use in QEC, but merely as used in a prescribed scheme. In principle, there can exist QEC schemes that are more or less sensitive to measurement imperfections than other. On the other hand,  a positive answer only indicates that the measurement is useful in principle, but other limitations, such as imperfect unitary control and slow data processing, could render the entire scheme useless. Despite these limitations, we believe that our scheme is of interest at  this early stage of information processing devices, and in particular for devices where distinct measurement schemes are currently being developped. 

The type of syndrome measurement that we have considered in this Article are of a special type that we call direct: the output signal at time $t$ is directly correlated with the value of the measured observable at that time (or a slightly earlier time). This type of measurement is to be distinguish from an indirect measurement where an ancillary system is coupled to the measured qubits and measured at a later time. For instance, a syndrome measurement can be realized by applying cnot gates to an ancillary qubit and then measuring that ancillary qubit in the computational basis. In a direct measurement, the signal history correlates with the noise history, so there is information to be gained by analyzing the entire signal rather than simply considering its average. In an indirect measurement, all the correlations between the error-corrected qubits and the ancillary qubit are established at the beginning (and on a very short time scale), so the signal history is uncorrelated to the noise history on the error-corrected qubit (but correlated instead to error history on the ancillary qubit). While some of the techniques we have proposed could be extended to an indirect measurement setting, it is not clear how to decouple the measurement imperfection from the imperfections arising in the unitary control in a measurement scheme that relies on unitary control.~\footnote{In the circuit QED architecture we have considered, the cavity is coupled to the qubit throughout the measurement process. Even though the cavity serves as an ancillary system, the measurement cannot be decomposed into a two stage process (unitary coupling followed by a measurement) and is thus of the direct type.}

Finally, another interesting error correcting code to study with this architecture is the Bacon-Shor code \cite{Pou05b,Bac05a,AC07a}. The main motivations for this code is that it is spatially local in a 2D architecture,  requires only two-qubit parity measurements, and can correct all type of errors. The complications arise because the parity measurements need to be in complementary basis, and our theoretical treatment would need to be generalized to subsystem codes \cite{KLP05a,Pou05b}.

\begin{acknowledgments}
We thank K.~Lalumi\`ere, M.~P.~da Silva and M.~Boissonneault for discussions and help with the numerical calculations. This work was partially funded by FQRNT, NSERC, the Alfred P. Sloan Foundation, and CIFAR. 
\end{acknowledgments}

\appendix

\section{Estimation of the average entanglement fidelity}
\label{sec:ave_entanglement_fidelity}

For a complete set of observables $A_j$ satisfying $\Tr(A_jA_k) = d\delta_{jk}$, for example Pauli operators where $d=2$, we can express the entanglement fidelity as
\begin{equation}\label{eq:HSP}
\begin{split}
F_e &=   \Tr(\Omega_{\cR_\cJ^\dagger} \Omega_{\cE_\cJ})\\
&=\frac 1d \sum_j \Tr(\Omega_{\cR_\cJ^\dagger} A_j) \Tr(\Omega_{\cE_\cJ} A_j).
\end{split}
\end{equation}
Let us denote by $\ket{\psi_\sigma^\tau} \in  \{\ket{0}, \ket{1}, \ket{0} \pm \ket{1}, \ket{0} \pm i{\ket 1}\}$ the eigenstates of the Pauli operators, i.e. $\sigma \ket{\psi_\sigma^\tau} = \tau \ket{\psi_\sigma^\tau}$ and similarly, let us denote by $\ket{\bar \psi_\sigma^\tau} \in \{\ket{\bar 0}, \ket{\bar 1}, \ket{\bar 0} \pm \ket{\bar 1}, \ket{\bar 0} \pm i\ket{\bar 1}\}$ the encoded versions. We can decompose the Pauli operator $A_j = B_k \otimes \sigma$ into a component $B_k$ on the $n$ physical qubits forming the QEC code and a Pauli matrix $\sigma$ acting on the reference qubit. This enables us to write
\BE
\Tr(\Omega_{\cE_\cJ} A_j) = \Tr\{B_k \cE_\cJ(\bar \Psi_\sigma^+)\} - \Tr\{B_k \cE_\cJ(\bar \Psi_\sigma^-)\}.
\label{eq:prod}
\EE
Combining with \eq{HSP}, we obtain
\BE
F_e = \sum_{k} \sum_\sigma \sum_{\tau = \pm} \tau \Tr(\Omega_{\cR_\cJ^\dagger} B_k\otimes\sigma) \Tr\{B_k \cE_\cJ(\bar \Psi_\sigma^\tau)\}.
\EE
Similarly, we can take a spectral decomposition of $B_k = P_k^+ - P_k^-$ to obtain 
\BE
F_e = \sum_{k} \sum_\sigma \sum_{\tau,\nu = \pm} \tau \nu \Tr(\Omega_{\cR_\cJ^\dagger} B_k\otimes\sigma) \Tr\{P_k^\nu \cE_\cJ(\bar \Psi_\sigma^\tau)\} 
\EE

The first term  $\Tr(\Omega_{\cR_\cJ^\dagger} B_k\otimes\sigma)$ of this expression can be evaluated numerically given knowledge of $\cR_\cJ$. Let us denote 
\BE
\Pr(\nu,\tau,k,\sigma) = \frac{\Tr\{P_k^\nu \cE_\cJ(\bar \Psi_\sigma^\tau)\} }{2\cdot 4^{n+1}}.
\EE
This can be interpreted as a probability distribution; it is positive and sums to 1. Moreover, one can efficiently sample from this distribution by choosing $k$ uniformly at random among the $4^n$ $n$-qubit Pauli operators; choosing $\sigma$ uniformly at random among the 4 single-qubit Pauli operators; $\tau$ uniformly at random between $\pm1$; and $\nu$ at random according to the probability $\Tr\{P_k^\nu \cE_\cJ(\bar \Psi_\sigma^\tau)\}$. These probabilities correspond exactly to probability with which the experiment described in Sec.~\ref{sec:delayed_tomography} generates the values $k$, $\sigma$, $\tau$, and $\nu$. 

Finally, rewriting
\BE
F_e =  4^{n+1}2 \sum_{k,\sigma,\tau,\nu} \tau \nu \Tr(\Omega_{\cR_\cJ^\dagger} B_k\otimes\sigma) \Pr(\nu,\tau,k,\sigma), \nonumber
\EE
we recognize \eq{F} as a Monte Carlo estimate of the entanglement fidelity $F_e$.

\end{document}